\documentclass[structabstract]{aa}
\usepackage{graphicx}
\usepackage{txfonts}

\def\ltsim{\raise 2pt \hbox {$<$} \kern-0.6em \lower 4pt \hbox {$\sim$}}
\def\gtsim{\raise 2pt \hbox {$>$} \kern-1.1em \lower 4pt \hbox {$\sim$}}

\def\ltapprox{\raise 2pt \hbox {$<$} \kern-1.1em \lower 5pt \hbox {$\approx
$}}
\def\gtapprox{\raise 2pt \hbox {$>$} \kern-1.1em \lower 5pt \hbox {$\approx
$}}

\begin{document}

\title{A radio minihalo in the extreme cool-core galaxy cluster RXCJ1504.1--0248}
\author{S.~Giacintucci\inst{1,}\inst{2},
M.~Markevitch\inst{2},
G.~Brunetti\inst{1},
R.~Cassano\inst{2,3},
\and
T.~Venturi\inst{1}}
\institute
{
INAF- Istituto di Radioastronomia, via Gobetti 101, I-40129, Bologna,
Italy
\and
Harvard-Smithsonian Centre for Astrophysics, 
60 Garden Street, Cambridge, MA 02138, USA,
\and
Dip. Astronomia, University of Bologna, via Ranzani 1,
I-40127 Bologna, Italy}


\titlerunning{A radio minihalo in RXC\,J1504.1--0248}
\authorrunning{Giacintucci et al.}

\abstract
{}
{We report the discovery of a radio minihalo in RXCJ1504.1-0248,
a massive galaxy cluster that has an extremely luminous
cool core. To date, only 9 radio minihalos are known, thus the
discovery of a new one, in one of the most luminous cool-core clusters,
provides important information on this peculiar class of
sources and sheds light on their origin.} 
{The diffuse radio source is detected using GMRT at 327 MHz and
  confirmed by pointed VLA data 
at 1.46 GHz. The minihalo has a radius of $\sim$140 kpc.
A Chandra gas temperature map shows that the minihalo emission fills the cluster cool
core and has some morphological similarities to it, as has been
previously observed for other minihalos.}
{The Chandra data reveal two subtle cold fronts in the cool core,
 likely created by sloshing of the core gas,  as observed in most
 cool-core clusters. Following previous work, we speculate that the origin 
of the minihalo is related to sloshing. Sloshing may result in 
particle acceleration by generating turbulence and/or amplifying the magnetic
field in the cool core, leading to the formation of a minihalo.}
{}

\maketitle
%

\section{Introduction}\label{sec:intro}

In some relaxed, cool-core clusters, the central radio galaxy
is surrounded by diffuse radio emission with $\ltsim$ 150-300 kpc
radius, with steep radio spectra ($\alpha >1$; $S_{\nu}
  \propto \nu^{-\alpha}$, where $S_{\nu}$ is the flux density at the
  frequency $\nu$) and low-surface brightness. These sources are
roundish in shape and very different from the lobes of the radio
galaxies often observed at the centre of cool-core clusters.
They are classified as radio minihalos (e.g., Ferrari et
al. 2008). Minihalos are rare; only nine cases have been
detected so far (e.g., Burns et al. 1992; Bacchi et al. 2003; Venturi
at al. 2007; Gitti et al. 2007; Govoni et al. 2009). Their physical
properties and origin are still poorly known. 

Minihalos cannot be explained by diffusion of relativistic
electrons from the central galaxy,  because the  
radiative lifetime of the electrons ($\sim 10^8$ yrs) is much shorter than the time 
needed for them to diffuse from the galaxy to the minihalo
radius. A possibility is given by reacceleration of pre-exisiting,
relativistic electrons in the intracluster medium (ICM) by turbulence
in the cooling flow (Gitti et al. 2002). 
Buoyant bubbles of relativistic plasma, inflated by the central AGN
and disrupted by the ICM motion, may provide the 
seed relativistic electrons that rapidly cool to energies where they
do not emit in observable radio band. Thus, minihalos may be the
ultimate depository of the relativistic matter ejected by the 
AGN over the cluster lifetime.  However, the strong gas inflows 
caused by the cooling 
appear unlikely with the present understanding of cooling flows (e.g.,
Peterson \& Fabian 2006). Thus, the origin of the turbulence needed to 
reaccelerate the electrons in minihalos is unclear. 
An interesting possibility is that it results from sloshing
of the core gas, frequently observed in the X-ray (e.g., Markevitch \&
Vikhlinin 2007). This was hinted at by the discovery of a
connection between minihalos and the sloshing cold fronts in two
clusters (Mazzotta \& Giacintucci 2008). The 
minihalos appear confined within these 
fronts, and the radio emission is spatially correlated with the spiral 
structure of the fronts. Thus, old relativistic electrons injected
over time by the central AGN may be spread out by gas sloshing and
reaccelerated by turbulence generated by this sloshing. Spiral flows
associated with sloshing may also amplify the magnetic fields
at the fronts (Keshet et al. 2010), which may amplify the synchrotron
emission. Hadronic collisions between thermal and cosmic ray protons 
can be alternative or additional sources for seed relativistic electrons
(e.g., Pfrommer \& Ensslin 2004).

In this Letter, we report the discovery of a radio minihalo at the
centre of the massive, cool-core cluster RXCJ1504.1-0248 
(hereafter RXCJ1504) at z=0.215 using archival {\it Giant Metrewave Radio Telescope} ({\it GMRT}) data and
1.46 GHz observations from the {\it Very Large Array} ({\it VLA}) archive.
The cluster exhibits extreme characteristics. With a total
X-ray mass of $1.8\times10^{15}$ M$_{\odot}$ within a radius of 3 Mpc
and a bolometric luminosity $L_{bol} = 4.3 \times 10^{45}$ erg
s$^{-1}$, this system is one of the most luminous clusters known 
(B{\"{o}}hringer et al. 2005). Its global X-ray temperature is 10.5 keV. 
The overall X-ray morphology is very compact and relaxed,
with a bright cool core with a cooling radius of $\sim$140 kpc,
in which the temperature drops below 5 keV. This extreme cool core accounts 
for more than 70$\%$ of the cluster total X-ray luminosity, 
making it one of the most luminous cool cores known (B{\"{o}}hringer et
al. 2005).
\\
We adopt 
H$_0$=70 km s$^{-1}$ Mpc$^{-1}$, 
$\Omega_m=0.3$ and $\Omega_{\Lambda}=0.7$. At the redshift of 
RXC J1504 (z=0.215), this gives 
$1^{\prime \prime}=3.49$ kpc.

%
%
\begin{figure*}
\centering
\hspace{-0.5cm}
\includegraphics[width=14.5cm]{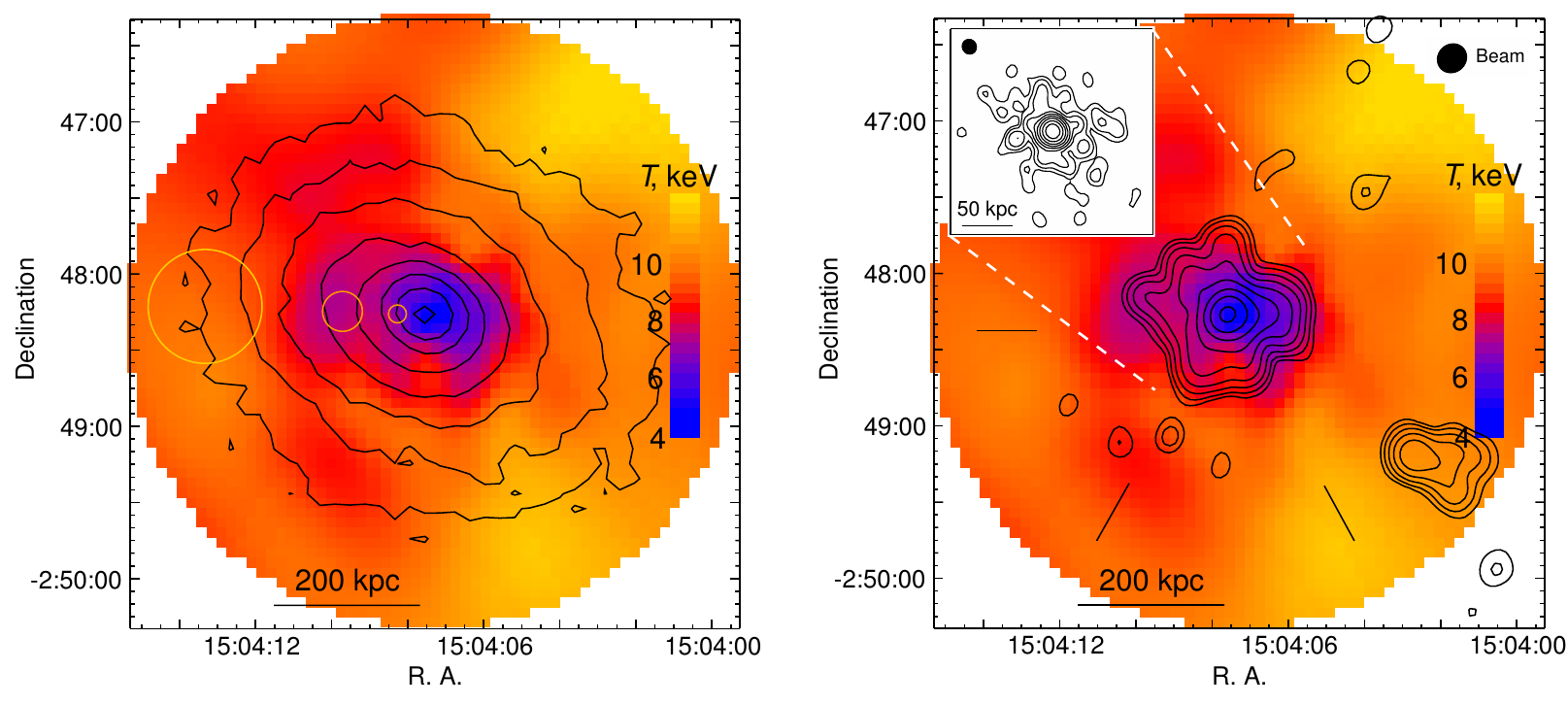}
\vspace{-1mm}
\caption{{\it Left:}\/ {\em Chandra}\/ gas temperature map (color);
 contours overlay the 0.8--4 keV {\em Chandra}\/ X-ray image smoothed with
  a $\sigma=2''$ Gaussian. Contours are spaced by a factor of 2 in surface
  brightness. The radii of three circles illustrate the variable smoothing
  width (Gaussian $\sigma$) used for the temperature map. The rms
  temperature uncertainties at their positions (from small to large radius)
  are 0.3 keV, 0.5 keV and 0.9 keV. {\it Right:}\/ {\it GMRT}\/ 327 MHz
  contours of the minihalo overlaid on the gas temperature map. The radio
  beam is $11.3^{\prime\prime} \times 10.4^{\prime\prime}$. Contours are
  spaced by a factor of 2, starting from $+3\sigma=0.3$ mJy beam$^{-1}$.
  Dashes show sectors used for radial profiles (Fig.\ 2).  {\it Inset:}\/
  {\it VLA}\/ 1.46 GHz pointed-observation image at resolution
  $4^{\prime\prime} \times 4^{\prime\prime}$. Contours start at 0.1 mJy
  beam$^{-1}$ and scale by factor of 2. Black ellipses show beam sizes of
  the two radio images.}
\label{fig:radio_x}
\end{figure*}
%
%

\section{Chandra data reduction}
We used archival {\it Chandra} data (ObsID 5793, ignoring another much
shorter observation in the archive) to obtain the X-ray image and
temperature map of the central region of RXCJ1504. The observation
was performed with ACIS-I with $\sim$40 ksec exposure. We cleaned the data
and modelled the detector background and instrumental spatial response as
described in Vikhlinin et al.\ (2005), using the latest {\it Chandra}
calibration (CalDB 4.3.0). The temperature map was derived as described in
Markevitch et al.\ (2000). Given the extremely bright cool core, we removed
the ACIS readout artifact following Markevitch et al.\ (2000). We excluded
point sources, produced background-subtracted and exposure-corrected images
in several energy bands (0.8-1.2-2.2-3.0-4.5-6.5-9.0 keV) and then smoothed
them by a Gaussian filter with the width dependent on the radius from the
brightness peak, in order to get good angular resolution at the peak and
good signal to noise ratio outside.  A temperature in each pixel of the map
was obtained by fitting values for each pixel of these images with a thermal
plasma model, with $N_H$ fixed to the Galactic value ($6\times 10^{20}$
cm$^{-2}$) and the metal abundance to 0.4 solar (an average value for a fit
to the central $r=2^{\prime}$ region).  The resulting temperature map in the
central $r=2^{\prime}$ is shown in Fig.\ 1 with X-ray contours overlaid
(left). The map clearly shows the cool core, where the temperature of the
gas drops to $\sim$4 keV. The size of the cool core agrees with the estimate
of the cooling radius ($\sim$140 kpc) by B{\"{o}}hringer et al.  (2005). The
eastern and western elongations of the cool core are significant at more
than $2\sigma$.

\begin{table*}[t]
\caption[]{Details of the radio observations of RXC J1504.1--0248 }
\begin{center}
\begin{tabular}{ccccccccc}
\hline\noalign{\smallskip}
Telescope & Project & Frequency & Bandwidth & Observation &
Integration  & FWHM, PA  &   rms & u-v range     \\  
  & & (MHz)      &      (MHz)   &      date        &  time (min)    &
  (full array, $^{\prime \prime} \times^{\prime \prime}$, $^{\circ}$)
  & ($\mu$Jy b$^{-1}$) & (k$\lambda$)\\
\noalign{\smallskip}
\hline\noalign{\smallskip}
GMRT & 05VKK01 & 327 &   32 &  2004 Apr 25 & 75 & 11.3$\times$10.4 ,
$-$44 &  100  & $\sim 0.1-26$\\ 
VLA--B &    AM938    & 1465 & 50 &  2009 Apr 6  &   160    &
4.26$\times$3.98, 29  &   35 & $\sim 0.8-55$\\
\hline
\end{tabular}
\end{center}
\label{tab:obs}
\end{table*}

\section{Radio observations and data reduction}\label{sec:obs}

RXCJ1504  was observed with the {\it GMRT} at 327
MHz as part of the {\it GMRT} Cluster Key Project (05VKK01).
Table 1 summarizes the details of the observation: frequency, bandwidth,
date, time on source, synthesized Full-Width Half-Maximum (FWHM)  and 
position angle (PA) of the full array,  rms
level (1$\sigma$)  at full resolution,  and u-v range.
The data were collected using the default spectral-line mode. Both
upper and lower side bands (USB and
LSB) were used,  providing a total observing bandwidth of 32
MHz. We calibrated the USB and LSB datasets 
individually using the NRAO Astronomical Image Processing System
(AIPS) package, following the procedure described in Giacintucci et
al. (2008). After bandpass calibration and a priori amplitude
calibration, a number of phase-only self-calibration cycles and 
imaging were carried out for each data set.  Multi-field
imaging was implemented in each step of the data reduction.
The USB and LSB data sets were then combined together to produce the final
image, which has a noise of 100 $\mu$Jy beam$^{-1}$ (Table \ref{tab:obs}).

We retrieved from the {\it VLA} archive the 1.46 GHz observation
of RXCJ1504 in B-array configuration (see Table 1 for details).
Calibration and data reduction were carried out in AIPS following the 
standard procedure (Fourier-transform, Clean and Restore). Phase-only 
self-calibration was applied to remove residual phase variations. The
final image has a noise of 35 $\mu$Jy beam$^{-1}$ (Table \ref{tab:obs}).
Average residual amplitude errors in the data are $\ltsim 5\%$ both at
327 MHz and 1.46 GHz.

\begin{figure}[t]
\centering
\includegraphics[angle=0,width=7.5cm]{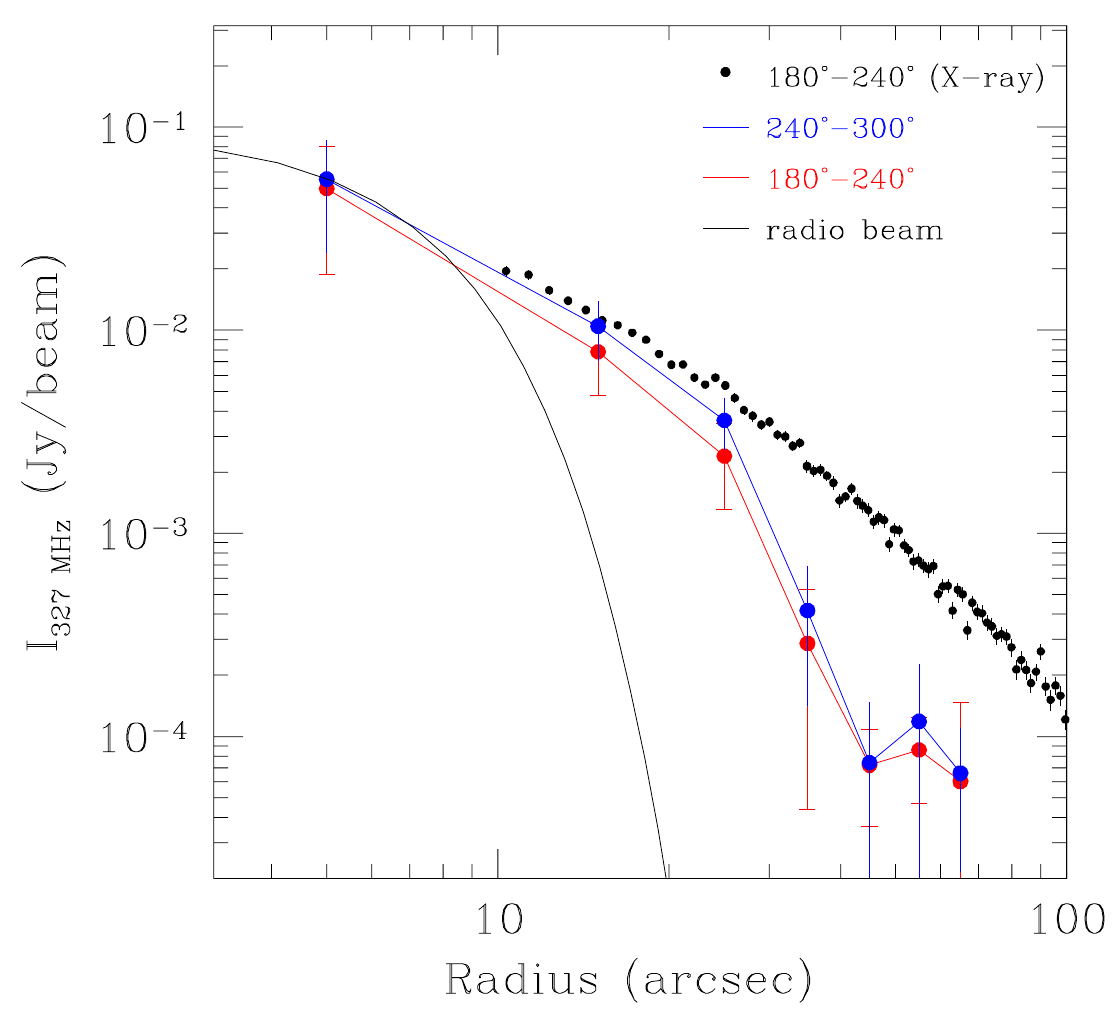}
\caption{327 MHz brightness profiles (blue and red) extracted 
in sectors shown in Fig.\ 1 (right), centered on the radio
peak. Comparison with the radio beam (black curve) clearly shows that 
the minihalo emission is extended. Black dots show the X-ray 
brightness profile in the same combined sector.}
\label{fig:front}
\end{figure}

\begin{figure}[t]
\centering
\includegraphics[angle=0,width=7cm]{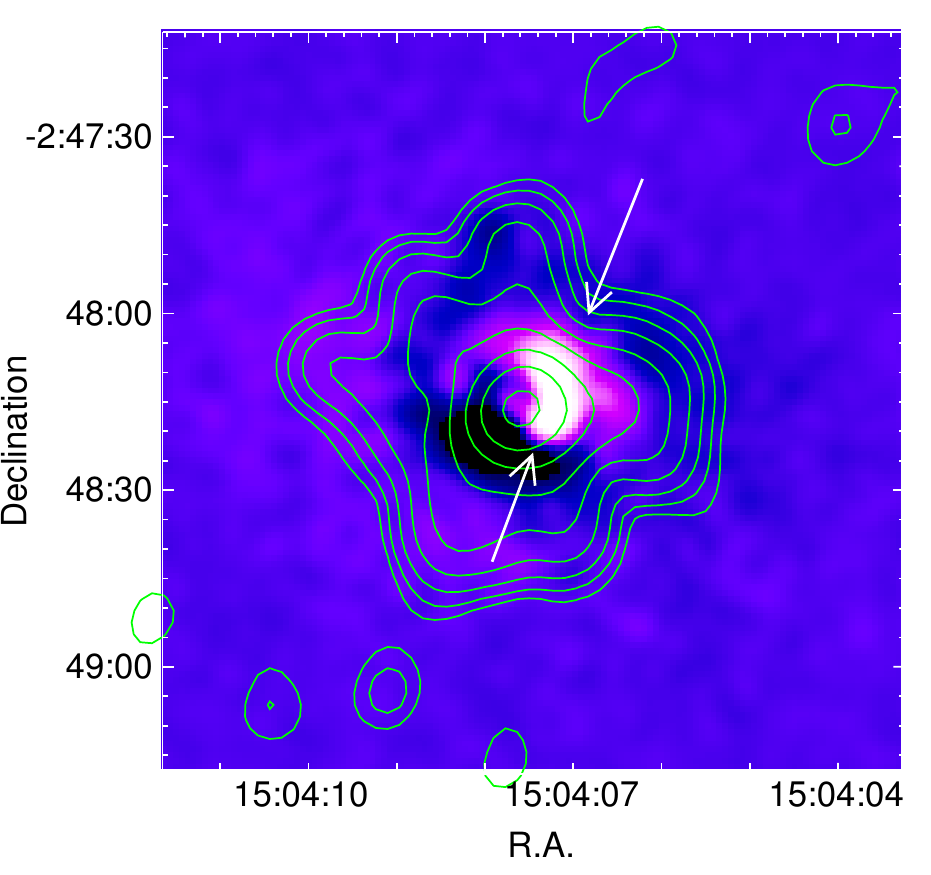}
\caption{A smoothed image of X-ray brightness residuals remaining after the
  subtraction of the best-fit elliptical $\beta$-model from the {\em
    Chandra}\/ image of the core region.  Two X-ray cold fronts are marked
  with arrows. Radio contours at 327 MHz are also shown (same as Fig.~1.).}
\label{fig:front}
\end{figure}

\section{The minihalo and the cool core}\label{sec:results}

In Fig.\ 1 (right), we overlay the {\it GMRT} 327 MHz contours on the
temperature map.  The central, unresolved radio component is associated with
the elliptical galaxy at the cluster center. The diffuse minihalo surrounds
this radio galaxy, extending for $\sim 40^{\prime \prime}$ ($\sim 140$ kpc)
in radius.  The {\it VLA} $4^{\prime \prime}$-resolution image at 1.46 GHz
is presented in the inset. The image shows diffuse emission on a slightly
smaller scale than the 327 MHz image (as expected for an interferometric
image with a higher resolution), while resolving more details near the
central compact source.  No clear radio structures, such as jets or tails,
connect the point source to the diffuse component, confirming that the
emission seen in the {\it GMRT} image is indeed a minihalo.

Fig.\ 1 shows that the minihalo is mostly confined within the cool core.
Some spatial radio features even coincide with similar structures of cool
gas in the temperature map. To verify this, we produced 327 MHz images with
lower resolution (not shown here) and found that the diffuse emission is
indeed confined within those radii.  Since the largest detectable structure
provided by the {\it GMRT} at 327 MHz is $\sim 30^{\prime}$, compared to the
$\sim 1.3^{\prime}$ size of the minihalo, we are confident that we are not
missing emission on a larger scale. Thus, the strong drop of the radio
brightness at the minihalo boundary is real. Fig.\ 2 shows radial profiles
of the radio emission extracted in two sectors in the south-east region of the core
(shown by black dashes in Fig.\ 1).  For comparison, we also show the X-ray
brightness profile extracted in the same combined sector (black dots).  
There is an abrupt drop in radio brightness at a radius $r > 25''$, while the 
X-ray profile continues smoothly beyond.

The 327 MHz flux density of the entire radio emission at the cluster center
is $S_{327 \rm \, MHz}$=215$\pm$11 mJy. The flux of the compact source
(obtained from a Gaussian fit in the full resolution image) accounts for
$\sim 44\%$ of the total flux, leaving 121$\pm$6 mJy to the minihalo. The
system is unresolved at the 45$^{\prime\prime}$-resolution of the 1.4 GHz
NRAO VLA Sky Survey (NVSS) image, where a total flux of 62 mJy is measured.
From the {\it VLA} image in the inset of Fig.\ 1, the contribution of the
central compact source at 1.4 GHz is 42$\pm$2 mJy.  Subtracting this from
the NVSS total flux, the 1.4 GHz flux density of the minihalo is 20$\pm$1
mJy.  The corresponding radio power is $P _{1.4 \rm \, GHz} =
3\times10^{24}$ W Hz$^{-1}$.  Using the fluxes at 327 MHz and 1.4 GHz, the
spectral index of the minihalo is $\alpha$\footnote{ $s\propto
  \nu^{-\alpha}$, where $S$ is the flux density at the frequency
  $\nu$.}=1.24, i.e., steep as typical for minihalos (e.g., Ferrari et al.
2008).  The central point source has a flatter spectrum ($\alpha$=0.56).
The different $u-v$ coverage of the 327 MHz and 1.4 GHz data sets 
at short baselines (Table 1) does not allow to derive a specral index
image of the minihalo.

The extent and 1.4 GHz power of the minihalo in RXCJ1504 
are in very good agreement with the size/power correlation found for
other minihalos by Cassano, Gitti \& Brunetti (2008).

\section{Gas sloshing in the cool core}

Inspection of the {\it Chandra} image reveals at least two subtle brightness
edges in the cool core. To better illustrate these edges, we fit a symmetric
elliptical $\beta$-model to the {\em Chandra}\/ image and subtracted it from
the real image. A smoothed residual image is shown in Fig.\ 3, with arrows
indicating the edges and 327 MHz contours overlaid. Comparison with the temperature map shows that these
are cold fronts, or contact discontinuities between two moving gas regions
with different temperatures and densities (Markevitch \& Vikhlinin 2007 and
references therein), which indicates sloshing of the cool dense gas in the
cluster core. A detailed check of their X-ray brightness profiles reveals
that these edges, while relatively sharp, are not well described by a simple
projected spherical gas density discontinuity, suggesting that sloshing
occurs not in the plane of the sky. Cold fronts are common in the cores of
otherwise relaxed clusters, and are believed to be created by sloshing of
the cool, dense core gas, possibly induced by a minor merger during the past
few Gyrs (e.g., Ascasibar \& Markevitch 2006).

\section{Discussion}
\label{sec:disc}

Despite extensive searches, less than ten clusters are known
to host a minihalo so far (Ferrari et al.\ 2008; Cassano,
Gitti \& Brunetti 2008; Govoni et al.\ 2009).  Thus, the
discovery of a new one in an extreme cool-core cluster
provides useful information on their origin and
the minihalo/cool-core connection.

Gitti et al.\ (2002) proposed that minihalos result from reacceleration of
pre-existing relativistic electrons in the ICM by turbulence in the cooling
flow.  More recenlty, based on the observed connection between radio
minihalos and X-ray cold fronts in several cool-core clusters, Mazzotta \&
Giacintucci (2008) proposed that, instead, the gas sloshing that creates
those cold fronts may be responsible for generating the turbulence needed to
reaccelerate the fossil relativistic electrons.  Indeed, high-resolution
numerical simulations show that sloshing can generate turbulence in the core
(e.g., Fujita et al.\ 2004; ZuHone et al.\ 2010 in preparation).  The radio
minihalo in RXCJ1504 appears confined to the cool core region,
suggesting a tight connection between the cool gas and the relativistic
plasma.  The X-ray cold fronts provide evidence that the gas in this cool
core is also sloshing, as observed in most cool cores (Markevitch \&
Vikhlinin 2007).  Thus, this mechanism may explain the mini-halo in this
cluster as well.

The acceleration timescale due to compressible turbulence is
$\tau_{acc} \sim 3 \times 10^6 ({{V_l}\over{c_s}})^{-4}
{{l/50}\over{c_s /1000}}$ yrs, where $l$ is the injection
scale (in kpc), $c_s$ is the sound speed and
$V_l$ is the velocity of turbulent eddies at the injection
scale (Brunetti \& Lazarian 2007, 2010).  Consequently,
$\tau_{acc} \sim {\rm few} \, 10^8$ yrs, necessary for 
reacceleration of GeV electrons, can be provided by this
mechanism if the energy of the turbulence, generated by
core sloshing (i.e., on a scale $l \approx 50-100$ kpc), is
$\sim 5-10 \%$ of the thermal energy.%
\footnote{We note that
the presence of other waves, such as Alfv\'en waves
generated at resonant scales, may imply even larger
fractions (compared to this case) of the turbulent energy
available for the reacceleration of relativistic particles,
making the required turbulent energy lower (e.g., Brunetti
\& Lazarian 2010).}
Recent X-ray spectral observations of A1835, a cool-core
cluster with a minihalo, placed an upper limit on the energy
of turbulence at $l \sim 30$ kpc scale in the cluster core of
$\sim 13\%$ of the thermal energy (Sanders et al.\ 2010), so
the required level of turbulence is allowed by current data.

In the context of the turbulent model, the high synchrotron
emissivity of minihalos implies an efficient supply of seed
relativistic particles to be reaccelerated in the emitting
volume, which may be provided by the central AGN (e.g.,
Cassano et al.\ 2007).  Because of a strong entropy decline
in the cool core, sloshing does not result in cool gas
spreading far outside of its equilibrium radius. Thus,
fossil relativistic electrons generated by the central AGN
should stay confined in the core, making the emission
process efficient.

Alternative models were also proposed to explain the origin of minihalos,
such as the secondary electron model (Pfrommer \& Ensslin 2004), in which
ultrarelativistic electrons originate from hadronic interactions of the
long-lived cosmic ray protons with thermal protons. A central AGN may also
supply the relativistic protons in the core.  However, what would create the
observed strong drop of the radio brightness at the boundary of the minhalo
in this model is not clear.  High-energy protons (with energies $\sim
30-100$ GeV) should diffuse on the 50--100 kpc scales on a short timescale,
$<1$ Gyr (e.g., Blasi, Gabici \& Brunetti 2007).  Assuming that the
resulting density of cosmic ray protons is proportional to the density of
thermal gas, $n_{\rm th}$, the synchrotron emissivity in secondary models
would scale as $j \propto n_{\rm th}^2 B^{1+\alpha} X / (B^2 + B_{\rm
  CMB}^2)$, where $B$ is the magnetic field intensity, $B_{\rm CMB}$ is the
equivalent field of the Cosmic Microwave Background photons and $X$ is the
ratio of the cosmic ray to thermal energy densities (Pfrommer \& Ensslin
2004). For a constant $B$, the synchrotron brightness should then follow the
X-ray brightness. However, while the radio minihalo has rather abrupt
boundaries, the X-ray brightness is smooth at those radii -- between
$r=30''$ and $40''$ from the centre (the interval containing the minihalo
boundary), the radio brightness declines by a factor of $\sim 7-8$, while
the X-ray brightness decreases only by a factor $\sim 1.5$ (Fig.\ 2). In the
framework of the secondary-electron model, this could be explained by a
highly amplified magnetic field within the cool core (possibly as a result
of gas sloshing, Keshet et al.\ 2010). The magnetic field strength should
then decrease to $B\ll B_{\rm CMB}$ outside the core.  Faraday rotation
measurements of cluster radio galaxies in systems similar to
RXCJ1504 can test this possibility.

\section{Conclusions}\label{sec:summ}

Using archival {\it GMRT} observations at 327 MHz and {\it VLA} 1.46
GHz data, we discovered a radio minihalo in RXCJ1504, with a radius of
$\sim$140 kpc.. The host cluster is very relaxed and possesses
one of the most luminous cool cores in the Universe, with 70$\%$ of the
total X-ray luminosity of the cluster coming from the core region. The {\it
Chandra} gas temperature map shows that the minihalo is confined to the
cool core. We found two cold fronts in the cool core,
suggesting that the core gas is sloshing. Such sloshing may
generate turbulence in the core, which in turn may reaccelerate relativistic
electrons, forming a minihalo.  Alternatively, the minihalo can be produced
by boosting of the synchrotron emission from an underlying population of
electrons (e.g., secondaries) if the magnetic field is strongly amplified in
the cool core.  In this case, the strong drop of the radio brightness at the
boundary of the cool core would imply that the magnetic field outside this
region is $B\ll B_{cmb}$.
\\
\\
{\it Acknowledgements.}  We thank the anonymous referee for constructive
comments and suggestions that improved this work. We thank 
P. Mazzotta for useful discussions. RC and GB thank
Harvard-Smithsonian Center for Astrophysics for hospitality.  {\em
  GMRT}\/ is run by the NCRA of the Tata
Institute of Fundamental Research. Financial support for this work was
partially provided by {\em Chandra}\/ grant AR0-11017X, NASA contract
NAS8-39073, by INAF under grants PRIN-INAF2007 and PRIN-INAF2008 and by
ASI-INAF under grant I/088/06/0.

\end{document}